\documentclass[pra,twocolumn,showpacs,floatfix,superscriptaddress]{revtex4-2}
\usepackage{graphicx}
\usepackage{epstopdf}
\usepackage{amsmath,amsthm,amssymb} 
\usepackage{color}
\usepackage{amsfonts}
\usepackage{subfigure}
\usepackage{bm}

\usepackage{epstopdf}
\usepackage{color}

\newcommand{\md}{\color{black}} 

\newcommand{\ket}[1]{\left| #1 \right>} 
\newcommand{\bra}[1]{\left< #1 \right|} 

\begin{document}
\title{High-fidelity quantum control via Autler-Townes splitting}
\author{Michele Delvecchio}
\affiliation{Dipartimento di Matematica, Fisica e Informatica, Universit\'a di Parma,  Parco Area delle Scienze 7/A, 43124 Parma, Italy}
\affiliation{Istituto Nazionale di Fisica Nucleare (INFN), Sezione Milano Bicocca, Gruppo di Parma, Parco Area delle Scienze 7/A, 43124 Parma, Italy}
\author{Teodora Kirova}
\email{teodora.kirova@lu.lv}
\affiliation{Institute of Atomic Physics and Spectroscopy, University of Latvia, Jelgavas street 4, Riga, LV-1004, Latvia}
\author{Ennio Arimondo}
\email{ennio.arimondo@unipi.it}
\affiliation{Dipartimento di Fisica, Universit\`a di Pisa, Largo Pontecorvo 3, 56127 Pisa, Italy}
\affiliation{Istituto Nazionale di Ottica - Consiglio Nazionale delle Ricerche,  Universit\`a di Pisa, Largo Pontecorvo 3, 56127 Pisa, Italy}
\author{Donatella Ciampini}
\email{donatella.ciampini@unipi.it}
\affiliation{Dipartimento di Fisica, Universit\`a di Pisa, Largo Pontecorvo 3, 56127 Pisa, Italy}
\author{Sandro Wimberger}
\email{sandromarcel.wimberger@unipr.it}
\affiliation{Dipartimento di Matematica, Fisica e Informatica, Universit\'a di Parma,  Parco Area delle Scienze 7/A, 43124 Parma, Italy}
\affiliation{Istituto Nazionale di Fisica Nucleare (INFN), Sezione Milano Bicocca, Gruppo di Parma, Parco Area delle Scienze 7/A, 43124 Parma, Italy}

\date{\today}


\begin{abstract}
	We propose quantum control protocols for the high-fidelity preparation of target-states in systems with Autler-Townes splitting. We investigate an approximated three-level system obtained from a four-level one by adiabatically eliminating a state that does not participate in the evolution. In our work we use linear, arctan and Roland-Cerf functions for transferring population between two eigenstates of the system obtaining a high fidelity for long evolution times. Additionally, in order to overcome the restriction given by the lifetimes of the experimental setup, we propose an accelerated adiabatic evolution with a shortcut to adiabaticity protocol, which allows us to reach fidelities close to one but much faster. 
 \end{abstract}
 
\maketitle
 
\section{Introduction}

Quantum control is based on the application of unitary transformations to simple or complex quantum systems to drive their evolution into a target quantum state~\cite{NielsenChuang:2010,DAlessandro:2007,WisemanMilburn:2009,GlaserWilhelm:2015}. Quantum control schemes should be classified on the basis of two elements, i.e., the experimental handle producing the quantum evolution, and the specific protocol describing the temporal modification of the system Hamiltonian. The detuning of a laser exciting the quantum system, for instance, a {\md two-level system}, is a convenient handle for preparing the initial quantum state and generating a Rabi evolution into the final state.  The protocol determines the time dependence of the handle, optimized to enhance fidelity, robustness and speed of the unitary transformation.

In a multilevel system with an intrinsic interaction between the internal states, the laser excitation may play a different indirect role.  The {\md non-resonant} control photons do not produce directly the quantum transfer: they alter the energy levels of the controlled system.  Following  the initial quantum preparation the final state transfer is driven by the internal Hamiltonian. Within this not-resonant approach the quantum control based on laser-induced potentials, i.e., on AC Stark shift, in some context known as Autler-Townes splitting, \cite{AutlerTownes:1955}, has  received a widespread interest, from atomic-molecular physics~\cite{GarrawaySuominen:1998,BrumerShapiro:2003,SussmanStolow:2006,SolaMalinovsky:2018}  to solid state systems~\cite{BoyleHopkinson:2009,AstapenkoYakovets:2015} and chemistry~\cite{StraniusBoerjesson:2018,EiznerKenaCohen:2019}.
Within the atomic-molecular physics area such an approach has been applied to  different processes as molecular alignment, photodissociation reactions, such as reviewed in~\cite{SolaMalinovsky:2018}. Our attention is focused on the quantum control of the two-electron  singlet-triplet  transitions in molecular systems, see e.g.  Refs. ~\cite{SolaMalinovsky:2006,GonzalezVazquezMalinovsky:2006,AhmedLyyra:2011,AhmedLyyra:2013,AhmedLyyra:2014,VindelSola:2013,FalgeSola:2014,Jamshidi-GhalehSahrai:2017}. 
The perturbed singlet-triplet states, arising from the level mixing due to the spin-orbit coupling, represent a gateway to gain access to the otherwise dark triplet states in these systems.  

In most cases, for instance, in ion strings~\cite{VindelSola:2013,FalgeSola:2014}, the Autler-Townes manipulation of the singlet-triplet transition  is produced by properly shaped short laser pulses. These pulses decouple the singlet and triplet electronic states, as well as minimize absorption to excited singlets and ionization of the absorbing species.  Refs.   \cite{AhmedLyyra:2013,AhmedLyyra:2014} introduce an alternative approach based on longer time scales, but leading to a more efficient singlet-triplet transfer.  Controlling the spin character of a spin-orbit coupled pair of levels, the Autler-Townes effect acts as an all-optical spin switch between singlet and triplet manifolds. In experiments on molecular states in lithium dimers this tuning produces an optical control of the singlet-triplet probability distribution.  Based on levels experiencing a weak singlet-triplet coupling  the Autler-Townes effect leads to a well controlled transfer to triplet states. In an equivalent quantum control experiment on perovskite nanocrystals exhibiting a spin-orbit induced multi-band structure,  the optical absorption is finely tuned using the Autler-Townes effect~\cite{YumotoKanemitsu:2021}. In addition Ref.~\cite{FengKwek:2018} applies an electrical control over the detuning energy of the quantum states of the two-electron singlet-triplet qubit across a chain of three quantum dots. 

Next to the experimental handle, the second control element is the specific time-dependent driving protocol. Detuning protocols such as the Landau-Zener-Majorana-St\"uckelberg tunneling (LZMS)~\cite{Landau:1932,Zener:1932,Stueckelberg:1932} have acquired a significant role  in different research areas. Relying on a linear dependence of a quantum handle it represents an easy to implement protocol for the preparation of the target quantum state. Both the asymptotic transition probability and its time dependence have  been examined extensively,  theoretically in Refs.~\cite{VitanovGarraway:1996,Vitanov:1999} and experimentally in cold atoms and solid state qubits~\cite{WilkinsonRaizen:1997,SillanpaaHakone:2006,ZenesiniArimondo:2009,KlingWeitz:2010,ZhouDu:2014,SunHan:2015}. Ref.~\cite{OlsonChen:2014} applied this protocol to a spin-orbit-coupled Bose-Einstein condensate. The LZMS tunneling is equivalent to the adiabatic passage of  light-induced potentials introduced in~\cite{GarrawaySuominen:1998} for controlling molecular reactions. Improvements in LZMS fidelity and speed  are obtained applying  a nonlinear temporal dependence of the control parameter~\cite{RolandCerf:2002,GaraninSchilling:2002,MalossiCiampini:2013a,StefanatosPaspalakis:2020}. In order to speed up the operation, more recently alternative superadiabatic protocols, referred to as "shortcuts to adiabaticity", have been introduced theoretically and experimentally as reviewed in~\cite{GueryMuga:2019}. 
  In order to avoid the additional requests on the parameter resources,  an effective instantaneous following of the adiabatic states  is achieved by introducing fast oscillations in the Hamilton parameters already present \cite{SuqingZhao:2005, PetiziolWimberger:2018}.
  
While the experiments of ~\cite{AhmedLyyra:2011,AhmedLyyra:2013,AhmedLyyra:2014} with lithium molecules are basically in a continuous-wave regime, the  present target is to modify the Autler-Townes quantum control in order to realize a time-dependent quantum transfer satisfying high-level requests for fidelity, robustness and speed. This scheme realizes a triggered and fast all-optical spin switch. The time dependence of the transfer protocol is based on the control of the Autler-Townes effect parameters. A similar time-dependent control of the Hamiltonian parameters can be applied to  other qubit systems experiencing a spin-orbit coupling. 

Our scheme based on the control of a probe laser frequency detuning, contains some features of the three-level Stimulated Raman Adiabatic Passage operation \cite{VitanovBergmann2017}. While the probe laser produces the excitation from the initial state to an intermediate excited one, the second step for the transfer to the target state is produced by the spin-orbit Hamiltonian. This second step not directly driven by a coupling laser, is controlled by the Autler-Townes effect of a second laser connecting the target state to a fourth level. Therefore an original feature of our theoretical analysis is the application of LZMS-like protocol to such an effective four-level system.
The conditions for the implementation of linear, non-linear and superadiabatic protocols are imposed on a reduced three-level system derived through an adiabatic elimination. The requests for such an elimination and the superadiabatic  realization in the full four-level system are carefully investigated. Even if our analysis is based on the experimental parameters of Ref. \cite{AhmedLyyra:2011,AhmedLyyra:2013}, the four-level LZMS protocol can be applied to a large variety of experimental schemes.

The paper is organized as follows: Section II introduces the level scheme and presents in detail the role of the spin-orbit coupling. In the following, the adiabatic protocol, the reduction of the four-level system to a three-level one, and the time-dependence of eigenvectors-eigenvalues of the full Hamitonian are presented. Section \ref{sec:protocols} examines the application of different adiabatic protocols to the four-level system and introduces the counterdiabatic approach as a tool to enhance the driving speed.
A short summary concludes our work. 

\begin{figure}[t]
  \centering
 \includegraphics[width= \linewidth]{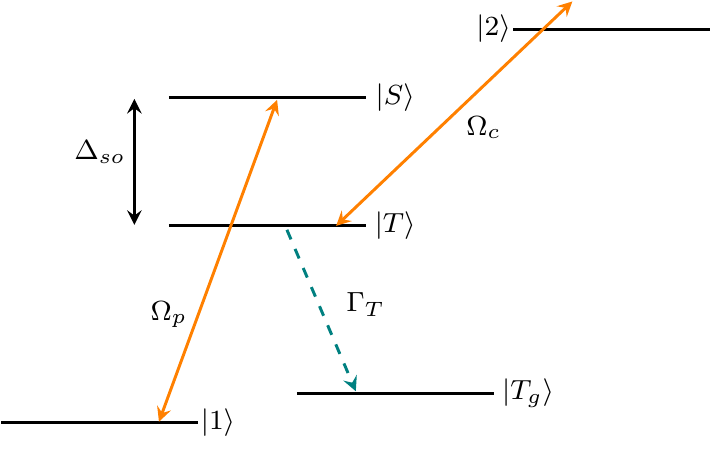}
  \caption{Four-level system with pump and coupling lasers, driving the transitions $|1\rangle \to |S\rangle$, with Rabi frequency $\Omega_p$ and detuning $\delta_p$, and $|T\rangle \to |2\rangle$, with Rabi frequency $\Omega_c$ and detuning $\delta_{c}$, respectively. A spontaneous decay with rate $\Gamma_{T}$ from the $|T\rangle$ state, represented by the dashed arrow, prepares the final state $|T_g\rangle$ of the all-optical switch. While in the experiment of Ref.~\cite{AhmedLyyra:2013} the coupling laser connects the $\ket{S}$ and $\ket{2}$ states, our equivalent choice allows an adiabatic elimination of the $\ket{2}$ state. The dressed states $|\pm\rangle$, arising due to the Autler-Townes mixing discussed below in Eq. \eqref{eq:dressed}, are shown for a blue detuned transition with $\delta_{c}>0$ as sketched in the figure. }
  \label{fig:LevelScheme}
\end{figure}

\section{Hamiltonian and driving schemes}
\label{sec:system}

In this section, we aim at providing the basis of the system setups that we used in our study. We first introduce the theoretical model of the four-level system we considered, then we show how and under what conditions it can be reduced to a three-level one, and finally we discuss the Autler-Townes effect in our context.

\subsection{Singlet-triplet Hamiltonian}

The singlet-triplet transfer experiment, as those in refs.~\cite{AhmedLyyra:2011,AhmedLyyra:2013,AhmedLyyra:2014}, are described by the four levels introduced in~\cite{KirovaSpano:2005} and presented in Fig. \ref{fig:LevelScheme}. {\md Their original structure for a lithium dimer is shown in Fig. 2 in Ref. \cite{AhmedLyyra:2013}. In particular, our initial ground state $|1\rangle$ would correspond in Li$_2$ to the rotational-vibrational state $X^1 \Sigma_g^+(\nu= 2, J = 22)$, and the spin-orbit coupled states $\ket{T}$ and $\ket{S}$ to $1^3 \Sigma_g^-(\nu= 1, J= 21, f)$ and $G^1\Pi_g(\nu=12,J=21,f)$, respectively. In the experiment reported in \cite{AhmedLyyra:2013} the first transition from the ground to the excited state is a two-photon one. Our target is to transfer the population from the singlet initial state $|1\rangle$ to the triplet excited state $|T\rangle$, and from the latter finally to the state $|T_g\rangle$ either automatically by spontaneous decay or by the application of an additional $\pi$-pulse.} The gateway key is the $(|S\rangle,|T\rangle)$ manifold of states with mixed singlet-triplet symmetries owing to a spin-orbit coupling. They are originated from the $(|S_0\rangle,|T_0\rangle)$  states corresponding to unperturbed singlet and triplet states with original energy separation $\Delta_0$. They experience a spin-orbit perturbation mixing with amplitude $V$ described by the Hamiltonian $H_{so}= V|S_0\rangle\langle T_0|+ H.c$, (assuming $\hbar=1$). The mixed states $|S\rangle$ and $|T\rangle$ are given  by
\begin{equation}
\label{eq:STeigens}
|S\rangle =\alpha |S_0\rangle- \beta |T_0\rangle, \qquad
|T\rangle =\beta |S_0\rangle+\alpha |T_0\rangle ,
\end{equation}
where the $(\alpha,\beta)$ coefficients are normalized to one, \textit{i.e.}, $|\alpha|^2+|\beta|^2=1$, while in the following these parameters are assumed to be real without loss of generality. The unperturbed energy separation $\Delta_0$ and the perturbation $V$ are linked to the effective energy splitting $\Delta_{so}$ of the $(|S\rangle,|T\rangle)$ levels and to the mixing coefficients by (see~\cite{KirovaSpano:2005}):
\begin{equation}
\label{eq:spinorbit}
\Delta_0= (\alpha^2-\beta^2)\Delta_{so} \qquad
V = \alpha\beta\Delta_{so}.
\end{equation}
The Lithium molecular states of Ref.~\cite{AhmedLyyra:2011} have mixing coefficients  $\alpha^2=0.87$ and $\beta^2=0.13$, and
 spin-orbit splitting $\Delta_{so} = 2\pi\cdot 0.75$ GHz equivalent to to $4.71$  ns$^{-1}$. In the following all parameters are given in the latter units, while the conversion into GHz is obtained dividing those by $2\pi$.

The access to the singlet-triplet manifold is provided by the pump ($p$) laser connecting the $\ket{1}$ singlet to the $|S\rangle$ state with detuning $\delta_p$.  The singlet  component of both  $|S\rangle, |T\rangle$ eigenstates of Eq.~\eqref{eq:STeigens} determines their excitation by the $p$ laser characterized by Rabi frequencies $\alpha\Omega_p$ and $\beta\Omega_p$, respectively. The effective laser excitation is controlled by the pump detuning from those states. Due to the mixed singlet-triplet character a fraction of molecules in the $|T\rangle$ excited state decays to the pure lower energy $|T_g\rangle$ triplet state. This process representing the basis of the optical pumping in triplet states for ultracold molecules is not very efficient for the rather small $\beta^2$ value reported above. 

The  quantum control of the singlet-triplet transfer introduced by~\cite{AhmedLyyra:2011,AhmedLyyra:2013} is based on the modification of the energy separation of the spin-orbit $|S\rangle, |T\rangle$ manifold. Such a modification is produced by a control  ($c$) off-resonant laser linking the $|T\rangle$ state to an additional $|2\rangle$ triplet one with detuning $\delta_c$.  The Rabi frequencies for the mixed manifold components are $-\beta\Omega_c$ and $\alpha\Omega_c$, respectively.

The temporal evolution of the optically driven levels $(|1\rangle,|S\rangle,|T\rangle, |2\rangle)$ is  described by a four-level Hamiltonian. For the $(|S\rangle,|T\rangle)$ basis we obtain
 \begin{equation} 
\label{eq:Hmatrix4level} 
H^{(4)}=
 \begin{pmatrix}
    \delta_p+\Delta_{so}& \alpha\Omega_p/2 &-\beta\Omega_p/2&0\\
    \alpha\Omega_p/2 & \Delta_{so} &0& \beta \Omega_c/2\\
    -\beta\Omega_p/2 & 0 &0&\alpha \Omega_c/2 \\
    0 & \beta\Omega_c/2 &\alpha \Omega_c/2  &-\delta_{c}
  \end{pmatrix},
\end{equation}
where we assume the zero energy at the position of the $|T\rangle$ state, and define $\delta_p=\omega_p-(E_S-E_1)$, $\delta_c=\omega_c-(E_2-E_T)$, with $E_i$ the energy of the states $(i=1,2,S,T)$. In the experiment of~\cite{AhmedLyyra:2011}, the Rabi frequencies were $\Omega_p=0.24$, $\Omega_c=3.8$, all in units of ns$^{-1}$.

While the lower $|1\rangle$ level  is stable and the $|2\rangle$ state is off-resonantly excited, the $|S\rangle,|T\rangle$ mixed levels suffer from spontaneous emission decay to levels outside the manifold of the above Hamiltonian, with decay rates of $\Gamma_{S}=0.06$ and $\Gamma_{T}=0.10$, both in ns$^{-1}$ for the experiments of ref.~\cite{AhmedLyyra:2011}. That experiment monitored the spontaneous decay to the  $|T_g\rangle$ triplet-ground state representing the final product of the all-optical switch, as denoted by the dashed line in Fig.~\ref{fig:LevelScheme}.
 
In our analysis we consider two regimes for $\delta_c$, supposed to be fixed during the protocol. The first is $\delta_{c} \approx 0$, in which the Autler-Townes effect must be considered. In the $|\delta_c| \gg \Omega_c$ second case, the problem can be reduced to a three-level model by adiabatically eliminating the state $\ket{2}$ from the beginning. 


\subsection{Reduced three-level System}
\label{sec:reduced_model}

For  $|\delta_{c}|\gg \Omega_c$, corresponding to a virtual excitation of the $|2\rangle$ state, the $\ket{2}$ state can be adiabatically eliminated. The corresponding reduced Hamiltonian reads
\begin{equation} 
\label{eq:Hmatrix3level} 
H^{(3)}=
  \begin{pmatrix}
    \delta_p +\Delta_{so}&\frac{\alpha\Omega_p}{2} & -\frac{\beta\Omega_p}{2}\\
    \frac{\alpha\Omega_p}{2} &\Delta_{so}+ \frac{\beta^2\Omega_c^2}{4\delta_{c}} &\frac{\alpha\beta\Omega_c^2}{4\delta_{c}} \\
    -\frac{\beta\Omega_p}{2} & \frac{\alpha\beta\Omega_c^2}{4\delta_{c}} &\frac{\alpha^2\Omega_c^2}{4\delta_{c}}\\
  \end{pmatrix}\,,
\end{equation}
now in the basis $\{\ket{1},\ket{S},\ket{T}\}$. 
Notice that within the effective energy of the state $|T\rangle$ the light-shift or ac shift $\delta_{ls}$ given by 
\begin{equation}
\label{eq:lightshift}
\delta_{ls}=\frac{\alpha^2\Omega_c^2}{4 \delta_{c}},
\end{equation} associated to the Autler-Townes process.  That term may produce a degeneracy between  the $|S\rangle$ and $|T\rangle$ 
for a blue detuning of the control laser and a proper choice of the control laser parameters. If we neglect the \textquotedblleft{}second order\textquotedblright{} Hamiltonian matrix elements proportional to $\beta^2, \alpha^2 \text{ and } \alpha\beta$, we obtain a matrix equivalent to Eq. (5) of Ref.~\cite{KirovaSpano:2005}. 
 
 \subsection{Autler-Townes effect}
\label{sec:AutlerTownesTheory} 

The driving of a two-level system by an intense resonant laser produces the Autler-Townes effect, i.e., the splitting of an optical transition~\cite{AutlerTownes:1955}. Within a dressed-atom  analysis~\cite{CohenTannoudji:1996} the degeneracy between the ground and the excited state is broken by the strong laser. The resulting energy separation is given by the Rabi frequency of the driving laser. Translated to our Fig.~\ref{fig:LevelScheme} level scheme, the Autler-Townes configuration of Refs.~\cite{AhmedLyyra:2011,AhmedLyyra:2013}  is created by the  coupling laser tuned at $\delta_c\approx 0$ modifying the energies of the states $|T\rangle$ and $|2\rangle$. That driving produces as eigenstates the following $|\pm\rangle$ superpositions: 
\begin{eqnarray}
\label{eq:dressed}
|+\rangle=&a_+|T\rangle+b_+|2\rangle, \nonumber \\
|-\rangle=&a_-|T\rangle+b_-|2\rangle,
\end{eqnarray}
with eigenvalues $\lambda_{\pm}$ given by
\begin{equation}
\label{eq:dressedeigen}
\lambda_\pm=-\frac{-\delta_c\pm\sqrt{\delta_c^2+\Omega_c^2}}{2}.
\end{equation}
and eigenstate coefficients given by
\begin{equation}
\frac{b_\pm}{a_\pm} = \frac{\delta_c \pm \sqrt{\delta_{c}^2 + \alpha^2\Omega_c^2}}{\alpha\Omega_c}.
\end{equation}
Because the Autler-Townes control is realized at low $\delta_c$ values, in this regime we usually cannot apply the adiabatic elimination reducing the four-level Hamiltonian of Eq.~\eqref{eq:Hmatrix4level} to the three-level of Eq.~\eqref{eq:Hmatrix3level}. 
However, in the next section we show that, using the states in Eq. \eqref{eq:dressed}, we obtain a matrix description in which $\ket{T}$ and $\ket{2}$ are not coupled. This will be useful for reaching a high fidelity in the Autler-Townes regime.
 
Because the energy separation between the states $|S\rangle$ and $|T\rangle$ may be modified by the Autler-Townes energy shift of the $|T\rangle$ state, their singlet-triplet coupling is also modified. In the experiment of Refs.~\cite{AhmedLyyra:2011,AhmedLyyra:2013} an increased triplet excitation reached approximately twice the value $\beta^2=0.13$ for the molecular states of interest. 

\subsection{Fidelity}

To measure the performance of the triplet transfer, we define the fidelity as the probability of being in the target state $\ket{T}$ at final time of the evolution $t=t_f$. In formula, 
\begin{equation}
\mathcal{F}(t_f)=\left|\langle\psi\left(t=t_f\right)|T\rangle\right|^2 \,
\label{eq:fidelity}
\end{equation} 
where $\psi(t=t_f)$ is the state of the system at the end of the protocol. As long as $t_f < \Gamma_T^{-1}$, our target state decays after the protocol into the experimentally relevant one $\ket{T_g}$, and the fidelities of both states will practically be identical. Alternatively, the population of the state $\ket{T}$ can be fast transferred to the final target $\ket{T_g}$ by a $\pi$ pulse. We will also use the infidelity $\mathcal{I} \equiv 1- \mathcal{F}$ if necessary for a better visualization.

\section{Laser driving protocols}
\label{sec:protocols}

This section introduces the adiabatic driving functions that we adopted in our schemes and then the protocol we chose to accelerate the adiabatic evolution of the mentioned driving functions.

\subsection{Adiabatic evolution}
\label{subsec:adiabatic}

\begin{figure}
	\centering	
	\includegraphics[width=\linewidth]{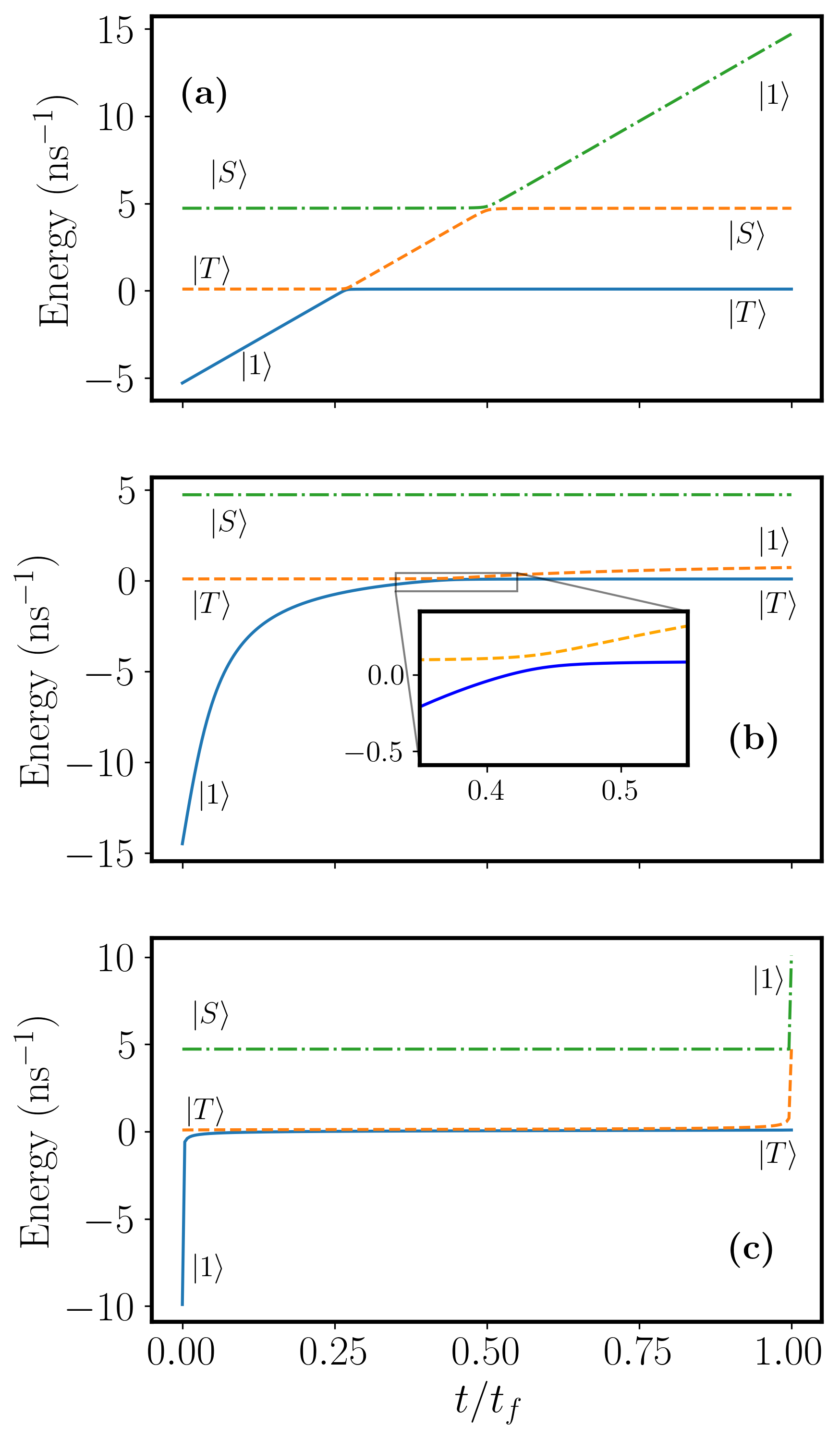}
	\caption{Instantaneous eigenvalues vs. rescaled time for the three sweep functions (a) linear; (b) arctan with $b=20$ and $c=19.2$ ns$^{-1}$; (c) Roland-Cerf with $d=4.68$ ns$^{-1}$.  In these plots $\delta_c=100$ ns$^{-1}$ leading to $\delta_{ls}=0.03$ ns$^{-1}$ for the light-shift defined in Eq.~\eqref{eq:lightshift}. On the left the position of the three initial eigenstates $\ket{1}, \ket{S}, \ket{T}$ and on the right their final energies are marked.  In (a) two avoided crossing are created, but we are only interested in the one between the states $\ket{1}$ and $\ket{T}$. In (b), thanks to the arctan function shape, only the avoided crossing of interest, observed in the inset, is generated. Also in (c) although with a different sweep function, a single avoided crossing is obtained. }
	\label{fig:reg1_eval}
\end{figure}

 Generalizing the idea of the LZMS problem~\cite{RolandCerf:2002,GaraninSchilling:2002,MalossiCiampini:2013a,PetiziolMannellaWimberger:2019,StefanatosPaspalakis:2020}, in order to realize the above goal we used specific functions such that the detuning $\delta_p$ is given by the following  dimensionless form
\begin{equation}
	\delta_p(\tau)=s f(\tau) \,,
\end{equation}
where $\tau=t/t_f$, $s$ is a scaling parameter and $f(\tau)$ a dimensionless function which we call \textit{sweep function}.  Exploiting the adiabatic theorem of quantum mechanics, which states that if the system changes slowly enough it follows the evolution of one of its initially prepared eigenstates adiabatically in time, we can design specific sweep functions in order to maximize the transfer to our target state.
In particular, we use three exemplary sweep functions: a \textit{linear} one from the classical LZMS problem, an \textit{arctan} function and the \textit{Roland-Cerf} \cite{RolandCerf:2002, PetiziolCM2019}. They explicitly read 
\begin{itemize}
	\item Linear \begin{equation}
		\delta_p(\tau) = a(2\tau-1) \,,
		\label{eq:lin_swp}
	\end{equation}
	\item Arctan \begin{equation}
		\delta_p(\tau)  = a\arctan(b \tau) - c ,
		\label{eq:arctan_swp}
	\end{equation}
	\item Roland-Cerf \begin{equation}
		\delta_p(\tau) =\frac{\beta\Omega_p(1-2\tau)}{2\sqrt{4\tau(1-\tau) +\left(\frac{\beta\Omega_p}{2a}\right)^2}} - d \,.
		\label{eq:roland_cerf}
	\end{equation}
\end{itemize}
\indent In our simulations we set $a=10$ ns$^{-1}$ producing the linear scan of Fig.\ref{fig:reg1_eval}(a).  The dimensionless parameter   $b$  of the arctan scan controls the shape of the $f(\tau)$  function and, consequently, the distance between the states $\ket{1}$ and $\ket{T}$ at the end of the protocol, as in Fig. \ref{fig:reg1_eval}(b). The Roland-Cerf function is based on the procedure described in \cite{PetiziolMannellaWimberger:2019} for the tangent protocol. The bias value  $d$ is optimized  to create the wanted avoided crossing, as in Fig. \ref{fig:reg1_eval}(c). An optimal choice of the parameters can improve the adiabatic evolution, but  may degrade the evolution with counterdiabatic pulses, as seen later on in Sec. \ref{sec:AT_results}.

For the three-level Hamiltonian in Eq. \eqref{eq:Hmatrix3level}, we compare the performance of the above sweep functions.  Fig.~\ref{fig:reg1_eval} reports the instantaneous eigenvalues of the Hamiltonian for the three functions when varying the detuning $\delta_p$ with time.  Because the energy of the triplet state $\ket{T}$ is lower than the $\ket{S}$ one, a scan of $\delta_p$ from a negative value with a positive slope reaches in time the $\ket{1} \to \ket{T}$ anti-crossing  before the $\ket{1} \to \ket{S}$ one, as for the linear sweep in (a). We are interested only in the first anticrossing where, adiabatically following the lowest eigenstate (blue/black solid line), a $\ket{1} \rightarrow \ket{T}$ population transfer  is realized. Therefore the linear sweep is terminated before reaching the $\ket{1} \to \ket{S}$ anticrossing.   In Fig. \ref{fig:reg1_eval}(b) or the arctan case, the coefficients $b$ and $c$ of Eq.~\eqref{eq:arctan_swp} governing the $\delta_p$ non-constant rate are chosen in order to avoid the second anticrossing. This generates temporal oscillations in the fidelity as shown later in the corresponding section. Also here an adiabatic following of the lowest eigenstate (blue/black solid line) produces the required population transfer.  In Fig. \ref{fig:reg1_eval}(c), such a transfer occurs in the instantaneous energies scheme for the optimized Roland-Cerf sweep, too.

 \subsection{Accelerated evolution}
 \label{sec:berry_theory}
 
The final time of transfer protocols must be compared to the spontaneous emission decay of the molecular states of interest. Therefore we accelerate the evolution in order to reach the required fidelity in a shorter time. However, an accelerated evolution typically generates unwanted non-adiabatic transitions between the eigenstates of the system, reducing the final fidelity. The approach introduced in \cite{Berry:2009,DemirplakRice:2003,FleischhauerBergmann:1999} drives the system in a perfectly adiabatic manner in an arbitrary short time. In the next sections, we show the result for the reduced three-level model, but the same procedure can be applied to the complete system as well.

Given a generic time-dependent system Hamiltonian 
\begin{equation}
	\hat{H}_0(t) = \sum_{k}E_k(t)\ket{k(t)}\bra{k(t)} \, ,
\end{equation}
where $E_k(t)$ and $\ket{k(t)}$ are, respectively, the instantaneous eigenvalues and eigenvectors of the system, the counter-adiabatic (CD) Hamiltonian $H_{CD}(t)$ is determined by:
\begin{equation}
\label{eq:HCD}
	H_{CD}(t) = i\hbar \sum_{n\neq k}\sum_{k}\frac{\ket{n(t)}\bra{n(t)}\partial_t H_0(t)\ket{k(t)}\bra{k(t)} }{E_k(t)-E_n(t)} \,.
\end{equation}
The system evolution is driven, thus, by the new Hamiltonian $\hat{H}(t)=\hat{H}_0 + H_{CD}(t)$. In this way, if the system starts in an eigenstate $\ket{k(t)}$, it remains in that instantaneous eigenstate for the entire evolution.  While for a two-level system the expression of $H_{CD}$ can be found analytically, for more levels it is generally more convenient to resort to numerical tools.  
{\md Identifying $\hat{H}_0$ with the matrices in Eqs. \eqref{eq:Hmatrix4level} and \eqref{eq:Hmatrix3level}, the Schr\"odinger equation with the total Hamiltonian $\hat{H}(t)$ is then numerically evolved using the function \textit{sesolve} of the python library \textit{qutip} \cite{JohanssonNori2013}, ensuring the convergence of the results.}

\begin{figure}[tb]
	\centering
	\includegraphics[width=\linewidth]{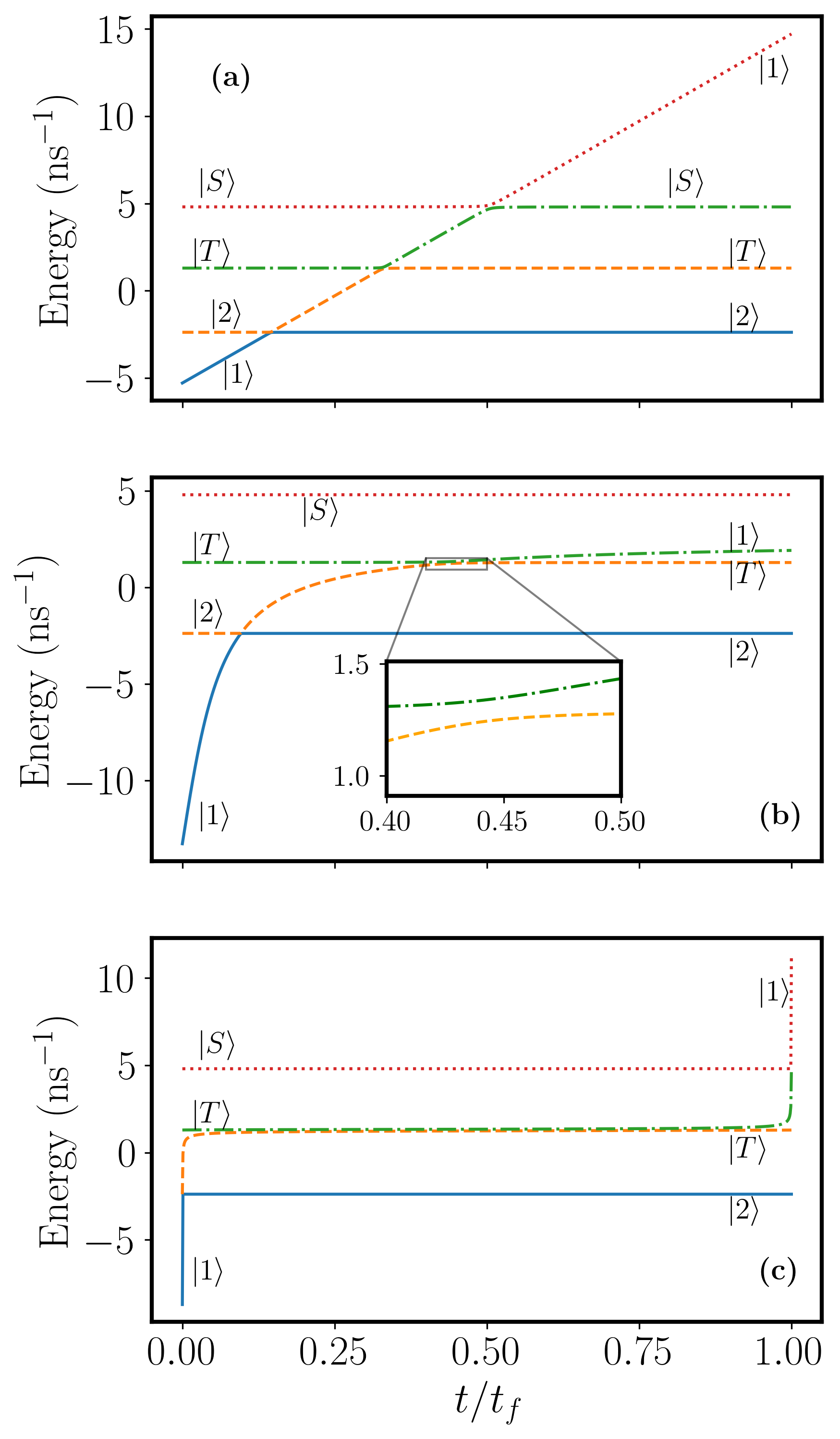}
	\caption{Instantaneous eigenvalues of the  four-level system vs the $\tau=t/t_f$ rescaled time at $\delta_c=1$ ns$^{-1}$ for the sweep functions: (a) linear; (b) arctan with $b=10$ and $c=18$ ns$^{-1}$; (c) Roland-Cerf with $d=3.41$ ns$^{-1}$.  The energies of the four states $\ket{1},\ket{S},\ket{T},\ket{2}$ evolve with time  from the initial to  final values. The impact of the final $\ket{1}$-$\ket{S}$ avoided crossing is handled as described in the caption of Fig.~\ref{fig:reg1_eval}. Note the presence of the $\ket{1}$-$\ket{2}$ additional avoided crossing with a small coupling strength. }
	\label{fig:4lev_eval}
\end{figure}

\section{Autler-Townes regime}
\label{sec:AT_results}

For the four-level system described by the Eq. \eqref{eq:Hmatrix4level}, we examine the population transfer between the states $\ket{T}$ and $\ket{1}$ at low $\delta_{c}$ values where the Autler-Townes effect plays a key role.  The instantaneous eigenvalues are depicted in Fig. \ref{fig:4lev_eval} for the protocols of Sec~\ref{subsec:adiabatic} with the parameters reported in the figure caption. The transfer of our interest is associated to the avoided crossing between the states $\ket{T}$ and $\ket{1}$. However  in all the protocols of Fig. \ref{fig:4lev_eval},  before reaching the level $\ket{T}$ the scanning reaches an additional $\ket{1}$-$\ket{2}$ avoided crossing. Even with a reduced strength this crossing would block the realization of a very high fidelity in the final $\ket{T}$  occupation. In order to bypass this loss, for both linear and arctan sweeps we split the $\delta_p$ evolution in two pieces. At first, using a short evolution time we scan diabatically the first avoided crossing.  Next we impose to $\delta_p$ the counter-diabatic evolution of Sec. \ref{sec:berry_theory}. The transition occurs at $\tau=t/t_f=0.2$. The temporal evolution of the Roland-Cerf sweep is automatically optimized for such fast-slow scanning. 

\begin{figure}[tb]
	\centering
	\includegraphics[width=.8\linewidth]{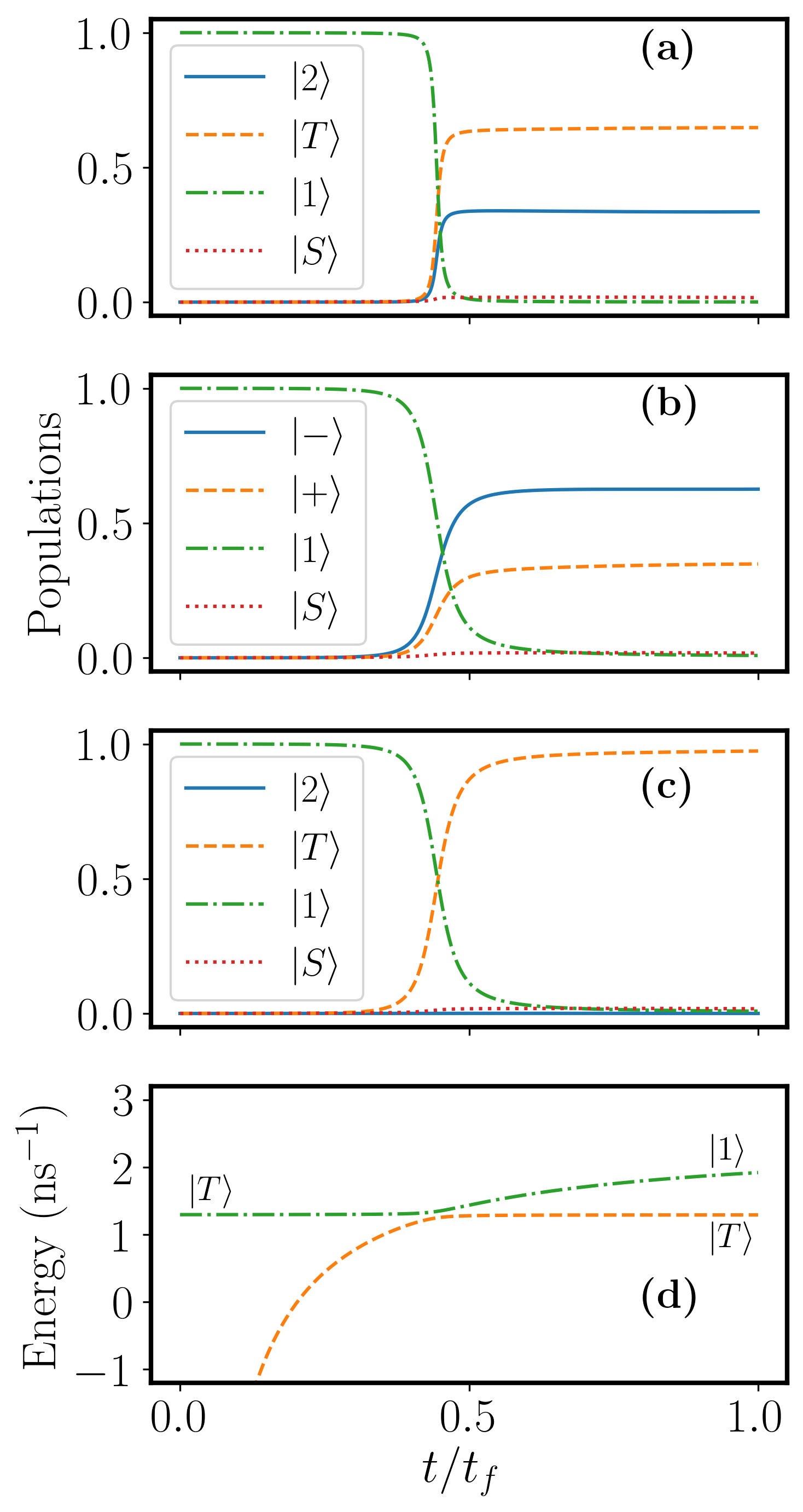}
	\caption{Populations of the four states obtained from the {\em different evolutions} in (a) with the four-level Hamiltonian of Eq. \eqref{eq:Hmatrix4level} in the original/bare basis, and  in (b) and (c) in the dressed representation introduced in Sec. \ref{sec:AT_results}. The populations are {\em measured} in (a) and (c) in the bare basis $\{\ket{T},\ket{2}\}$ and in (b) in the dressed basis $\{\ket{\pm}\}$. {\md The population of $\ket{T}$ can then be transferred to $\ket{T_g}$ either by spontaneous decay or by applying an additional $\pi$ pulse.} All evolutions are performed using the arctan sweep together with the shortcut to adiabaticity protocol presented in Sec. \ref{sec:berry_theory} for $\tau = t/t_f > 0.2$. We choose a very fast evolution with {$t_f=1$ ns} in order to diabatically pass through the first avoided crossing. At $t_f=1$ ns, the measured $\ket{T}$ population is 97\% in (c), much higher than for the different evolution in (a). {\md (d) shows a zoom of Fig. \ref{fig:4lev_eval}(b) into the interested avoided crossing. Please notice that the population transfer in (a),(b) and (c) occurs at the position 
	of the avoided crossing in (d).}}
	\label{fig:AT_pops1}
\end{figure}

As an example of the fidelity results, Fig. \ref{fig:AT_pops1}(a) shows the populations of the four levels, of the original basis, in the case of the arctan protocol.  The $\ket{T}$ state fidelity reaches a final value around sixty-five percent.  However, even with a perfectly adiabatic evolution, a large part of the final population, $\approx 35\%$ in this case, occupies the state $\ket{2}$. This result appears because the counter-adiabatic protocol imposes an evolution linking initial and final superposition of states. For the present case, the Autler-Townes effect produces a close degeneracy between the states $\ket{1}$ and $\ket{+}$. Figure \ref{fig:AT_pops1}(b) represents the populations of the dressed states obtained from the evolution of the dressed basis $\ket{\pm}$.

Therefore, we have applied the counter-diabatic evolution to the new four level system where the  $\ket{T},\ket{2}$ states are now replaced by the dressed basis $\ket{\pm}$ of Eqs.~\eqref{eq:dressed}.
Figure \ref{fig:AT_pops1}(b) shows the populations of the four states $\ket{+}, \ket{S}, \ket{1}, \ket{-}$ obtained from the evolution of the CD Hamiltonian in the dressed representation. Figure \ref{fig:AT_pops1}(c) reports the populations of the four states $\ket{T}, \ket{S}, \ket{1}, \ket{2}$ derived from that evolution by performing a projective measurement of the occupations of the states $\ket{T}$ and $\ket{2}$.  The states $\ket{\pm}$ are superpositions of $\ket{T} , \ket{2}$. Thus, the measured  $\ket{T}$ population contains contributions  from both dressed states leading to interference effects. The  CD dressed Hamiltonian with a maximized fidelity produces a final 97\% occupation of $\ket{T}$,  higher than that in the case (a). This population can be transferred by a $\pi$ pulse on a short time scale to the final experimental target $\ket{T_g}$.  The spontaneous emission $\ket{T}\to\ket{T_g}$ with $1/\Gamma_T \gg t_f$ represents an alternative for the accumulation into the target state.  

\section{Three-level performances}
\label{sec:reduced_results}
In this section, we return to the three-level system introduced above by Eq. \eqref{eq:Hmatrix3level}. We first analyze the fidelities of the driving protocols when no CD correction is applied, then we accelerate the evolution introducing the CD term presented in Sec. \ref{sec:berry_theory}.

\begin{figure}
	\centering
	\includegraphics[width=\linewidth]{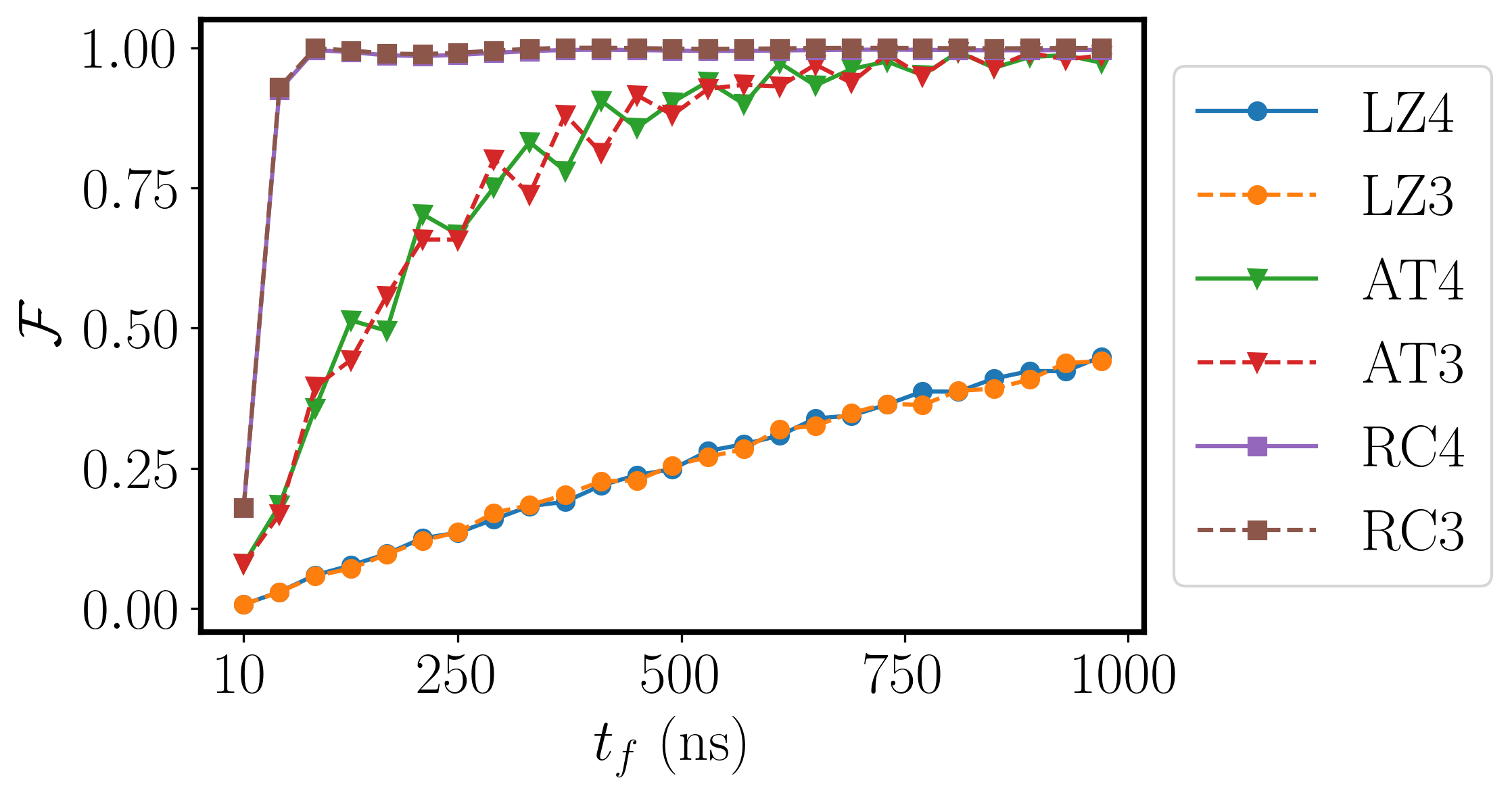}
	\caption{Comparison between the fidelity, at $t=t_f$, as defined in Eq. \eqref{eq:fidelity}, of the three adiabatic protocols denoted as LZ, AT and RC, at  $\delta_{c}=30$ ns$^{-1}$. The dashed lines represent the reduced three-level system, while the solid lines describe the complete four-level system. The two systems are identified by the numbers $3$ and $4$ in the legends. For such a large $\delta_c$, the four and three-level systems produce very similar fidelities.}
	\label{fig:fid_comparison3}
\end{figure}

 \subsection{Linear, arctan and Roland-Cerf protocols}
\label{sec:4lev}

\begin{figure}
	\centering
	\includegraphics[width=.96\linewidth]{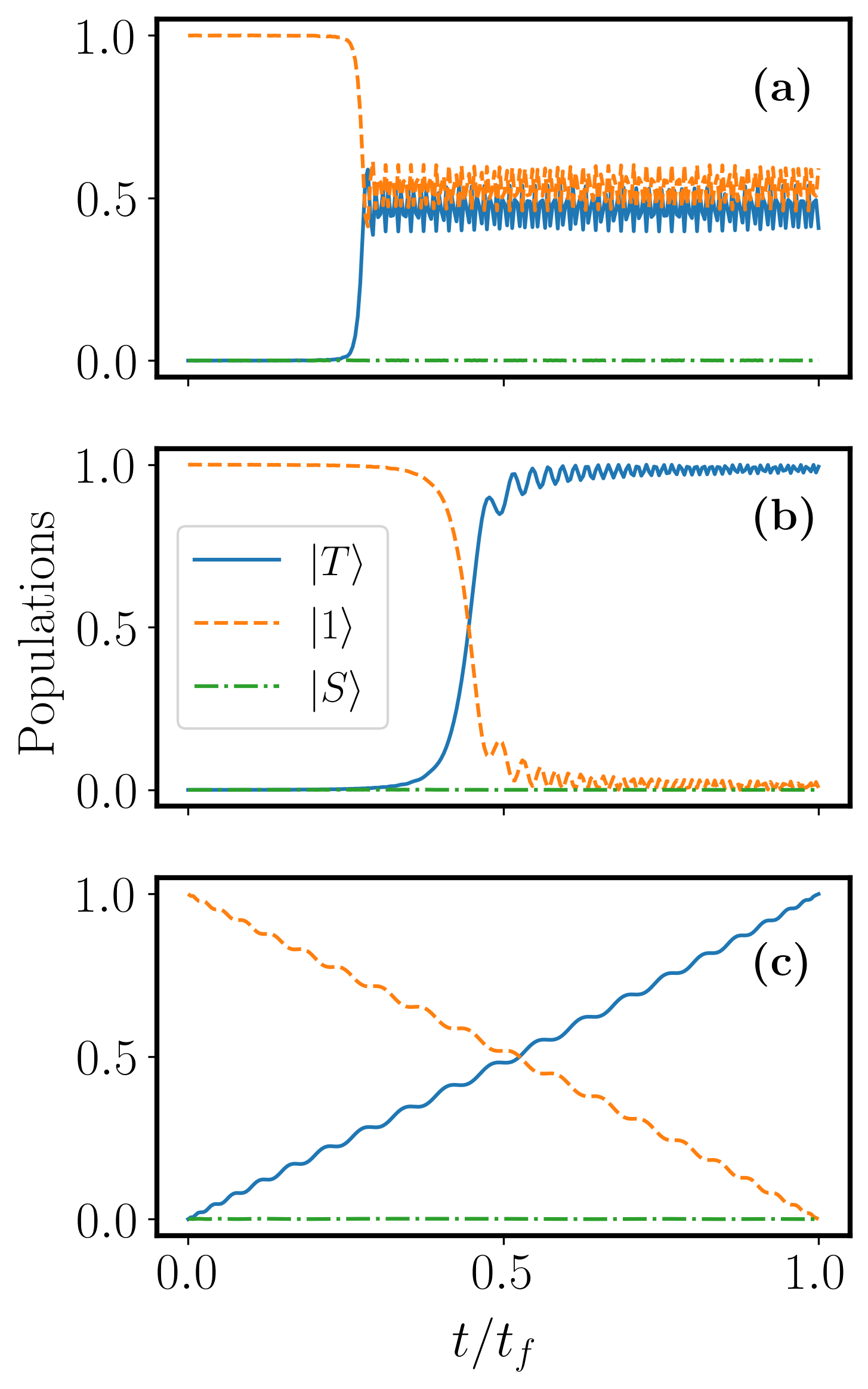}
	\caption{Populations of the three levels at $t_f=1000$ ns and $\delta_c=30$ ns$^{-1}$, for the (a) \textit{linear}, (b) \textit{arctan} and (c) \textit{Roland-Cerf} sweep functions. {\md The populations in all the three cases are represented by the orange dashed line for the state $\ket{1}$, the blue solid line for $\ket{T}$ and green dash-dot line for the state $\ket{S}$. The legend in (b) is valid also for (a) and (c).} For the same evolution time, the latter two driving protocols produce better results.  The population of the state $\ket{T}$ represents the fidelity as a function of the rescaled time. Oscillations in (a) are caused by the fact that the linear protocol is terminated before reaching the level $\ket{S}$ (at $t/t_f = 0.3$ in this case) and, therefore, it assumes a constant shape for the rest of the evolution time.}
	\label{fig:atan_pop}
\end{figure}

Figure \ref{fig:fid_comparison3} reports a comparison in fidelity of the three protocols presented in Sec. \ref{sec:protocols}, without the counter-diabatic term and in the regime of large values of $\delta_{c}$, such that the $|2\rangle$ can be eliminated and the system is effectively reduced to a three-level one.
The well designed Roland-Cerf driving reaches a better performance with $\mathcal{F}=93 \%$ at $\tau_f=50$ ns for $\delta_c=30$ ns$^{-1}$, as shown in Fig. \ref{fig:fid_comparison3} by the dashed line with square symbols. Good fidelities are reached at longer final times with the arctan protocol. However, in this case the population of the level $\ket{T}$ experiences  temporal oscillations, as shown Fig. \ref{fig:atan_pop}(b). Similar oscillations are observed also in Fig~ \ref{fig:fid_comparison3}. The amplitude of these oscillations is reduced  using  larger $t_f$ values. Lower fidelities are obtained with the linear protocol, with large oscillations in the populations. These are generated by changing the linear protocol to a constant after $t/t_f = 0.3$. As above, this change is necessary if one wants to avoid to create an anticrossing with $\ket{S}$ and transfer, in this way, part of the population to it. 

\subsection{Counter-diabatic evolution}

\begin{figure}
	\centering
 	\includegraphics[width=.9\linewidth]{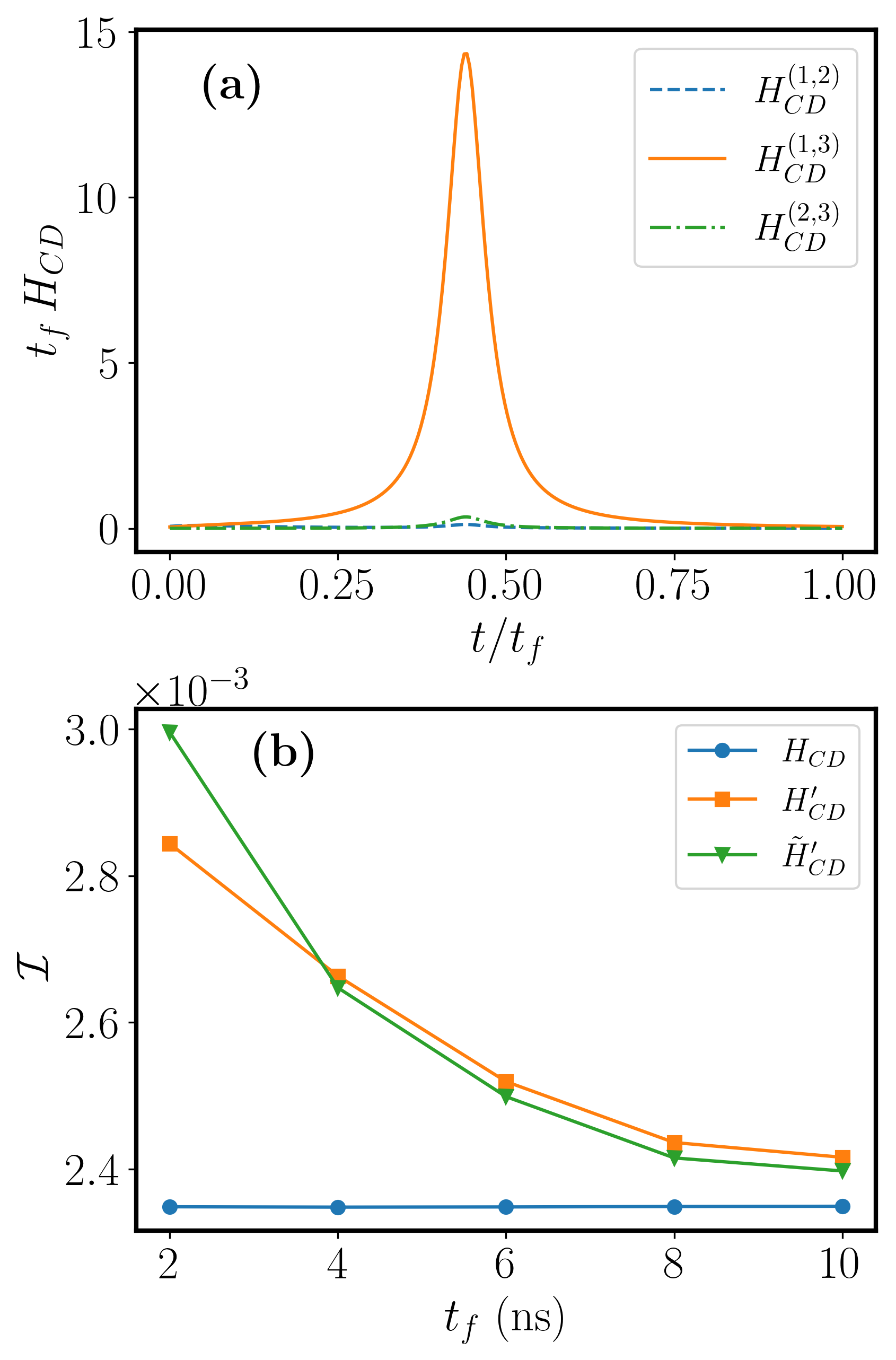}
	\caption{In (a) imaginary parts of the $H_{CD}^{(i,j)}$ elements  with $i,j=1,2,3$ for the arctan protocol, with $H_{CD}^{(1,3)}$ the largest required pulse. Parameters  $\delta_{c}=30$ ns$^{-1}$, $t_f=1$ ns, $c=19.2$ ns$^{-1}$, $b=20$. In (b) infidelities, at the $10^{-3}$ level, for different CD corrections. Blue dots: for full $H_{CD}$ elements;  orange squares: eliminating the $H_{CD}^{(2,3)}$ correction only; green triangles: eliminating both $H_{CD}^{(2,3)}$ and  $H_{CD}^{(1,2)}$ corrections. }
	\label{fig:berry_pulses}
\end{figure}

We report here  the performance of the arctan  $\ket{1} \to \ket{T}$  protocol as corrected by the Berry counter-diabatic Hamiltonian  of Eq.~\eqref{eq:HCD}.  For the three-level system  three additional pulses are required in order to eliminate all the non-adiabatic transitions. The imaginary part of those $H_{CD}^{(i,j)}$ matrix elements  are shown in Fig. \ref{fig:berry_pulses}(a). Applying these correction pulses, we obtain infidelities of the order of $10^{-3}$ as in the blue solid line in Fig. \ref{fig:berry_pulses}(b) for an arbitrary evolution time.\\ 
\indent The application of the Berry correction  pulses is  costly in terms of energy and of new matrix elements in the Hamiltonian.  Towards a simplified experimental realization,  the number of $H_{CD}$  corrections may be reduced to the most relevant ones, as analyzed in \cite{PetiziolMannellaWimberger:2019}.  Fig.~\ref{fig:berry_pulses}(a) evidences that the $H_{CD}^{(1,3)}$ correcting peak is greater than the remaining ones.  Fig. \ref{fig:berry_pulses}(b) compares the infidelities applying the full  $H_{CD}$, and eliminating either one or both the small amplitude corrections. The elimination of the two terms in the CD Hamiltonian produces only a maximum change at the $5\cdot 10^{-4}$ level, but simplifies a lot the experimental realizations. Therefore the single $H_{CD}^{(1,3)}$ implementation produces very high fidelities on timescales much shorter than without the acceleration, bringing experimental times down to the coherence time of the system. A similar reduction of the effect number of matrix elements in  $H_{CD}$ is possible in the protocol shown in Sec. \ref{sec:AT_results}, but will not be shown here in full detail.

 \section{Conclusions}
 
We have explored the Autler-Townes control of spin-orbit coupling in an original four-level system with the target of a large transfer efficiency to a final triplet state.  We extended the flexibility explored by~\cite{Jamshidi-GhalehSahrai:2017} for the nonlinear response in such a system to population transfer. Our work confirms the interesting feature of a probe-coupling excitation laser scheme, assisted by the intermediate spin-orbit coupling. However the present system is not equivalent to a STIRAP configuration \cite{VitanovBergmann2017} {\md which realizes a population transfer in, e.g., a three-level system without populating the intermediate state. Our target, instead,} is to populate the final triplet state modifying the spin-orbit coupling of the intermediate and final states. The quantum control is based on the Autler-Townes produced a modified energy separation between the singlet and triplet states. The detuning and the intensity of the control laser represent the key parameters of the control operation.

We have applied the counter-diabatic control in order to speed up the transfer process with a high transfer fidelity. Our population transfers are all larger than those reached in the experiments of \cite{AhmedLyyra:2011}. The use of accelerated adiabatic driving produces the required transfer in a time short compared to the decay of the molecular states of interest. These features make our protocols very relevant for future experiments.

For the present population transfer the coupling between initial and final state is based on an intermediate singlet-triplet coupling Hamiltonian. However the present analysis may be applied as well, e.g., to configurations where that coupling is produced by an applied magnetic field, static or oscillating, or by a two-photon laser interaction. Thus, our quantum transfer may be useful to a large class of quantum simulations. In heavy alkali dimer molecules such as Rb$_2$ and Cs$_2$, the spin-orbit interaction is much stronger than for the lighter ones. Such perturbed levels are used as intermediate levels in the transfer of cold molecules formed at long range in the triplet  state to deeply bound levels of the singlet ground state. Our control approaches could be applied also to those systems.
 
\begin{acknowledgments}
TK acknowledges support from the Visiting Fellow Program of the University of Pisa. TK and EA thank A. H. Ahmed and A. M. Lyyra for extended stimulating discussions. We are grateful to Francesco Petiziol for comments on the manuscript.
\end{acknowledgments}

%

\end{document}